# Metamorphic Testing:
# A New Approach for Generating Next Test Cases [†]

T. Y. Chen

Department of Computer Science, The University of Melbourne

Parkville 3052, Australia

(tyc@cs.mu.oz.au)

S. C. Cheung[§]

Department of Computer Science, The Hong Kong University of Science and Technology

Clear Water Bay, Kowloon, Hong Kong

(scc@cs.ust.hk)

S. M. Yiu

Department of Computer Science, The University of Hong Kong

Pokfulam Road, Hong Kong

(smyiu@cs.hku.hk)

**Abstract**

In software testing, a set of test cases is constructed according to some predefined selection criteria. The software is then examined against these test cases. Three interesting observations have been made on the current artifacts of software testing. Firstly, an *error-revealing* test case is considered useful while a *successful* test case which does not reveal software errors is usually not further investigated. Whether these successful test cases still contain useful information for revealing software errors has not been properly studied. Secondly, no matter how extensive the testing has been conducted in the development phase, errors may still exist in the software [5]. These errors, if left undetected, may eventually cause damage to the production system. The study of techniques for uncovering software errors in the production phase is seldom addressed in the literature. Thirdly, as indicated by Weyuker in [6], the availability of test oracles is pragmatically unattainable in most situations. However, the availability of test oracles is generally assumed in conventional software testing techniques. In this paper, we propose a novel test case selection technique that derives new test cases from the successful ones. The selection aims at revealing software errors that are possibly left undetected in successful test cases which may be generated using some existing strategies. As such, the proposed technique augments the effectiveness of existing test selection strategies. The technique also helps uncover software errors in the production phase and can be used in the absence of test oracles.

---

[†] This project was partially supported by a grant from the Australian Research Council and the Hong Kong Research Grant Council.
[§] Corresponding Author

# 1 Introduction

Although errors exist in most software systems, they cannot be tolerated in general. This is particularly the case if software systems are deployed for safety-critical applications. Improving software reliability has been a major issue in software engineering. *Software Testing* is by far the most widely used approach to examine the correctness of a piece of software before it is released for production. In software testing, a piece of software is tested against a set of test cases generated by a predefined test selection strategy. A test case is *error-revealing* if it detects a software error; otherwise it is called *successful* [4]. Error-revealing test cases are generally regarded to be more informative than the successful ones as they demonstrate the existence of software errors. Correction of such errors improves the software reliability.

No matter how extensive testing has been conducted in the *testing phase* during software development, errors may still exist in a software system after it has been released for production [5]. Due to the massive volume of production data, outputs of a system in the production phase are not normally verified. If these errors remain undetected in the production phase, they can have catastrophic consequences. Hence, new techniques for identifying program errors in the production phase should be developed. In view of this, Blum [1, 2] recently introduced the concept of *program checking*. A *program checker* is another program specially written to verify the output of a given program. This checker can be "embedded" in a system for automatic output verification in both the testing and production phases. A program output is definitely incorrect if it is rejected by the checker. However, acceptance of the output does not guarantee its correctness. In other words, not all incorrect outputs can be identified and there may still be program errors which are not revealed by the current input. Obviously, more should be done in enhancing the confidence of the software in the production phase.

Software testing techniques generally assume the availability of a test oracle. However, as indicated by Weyuker [6], this assumption may not hold in practice. For example, it is difficult to check if a path in a non-trivial graph is the shortest. This is particularly the case for applications in their production phases. It is therefore useful to have a technique which does not assume the availability of such oracles. In this paper, a novel technique for test cases selection is proposed which constructs new test cases from successful test cases, aiming at revealing further software errors. The technique also helps uncover software errors in the production phase and can be used in the absence of test oracles. As our method is applicable to applications in both development and production phases, for simplicity, the inputs in the production phase will hereafter be referred to as test cases. A test case together with its test result will be referred to as an input-output pair. In our approach, the generation of new test cases is based on the input-output pairs of previous test cases (in particular on the successful ones) and the types of errors usually associated with that particular type of applications. In other words, we always assume that there are errors inside the software in spite of the apparently correct output produced from the current input. The approach is considered *fault-based* as new test cases aim at uncovering specific errors which left undetected in previous successful test cases. We call this approach *Metamorphic Testing* because new test cases are evolved from the old ones that are normally selected according to some existing selection criteria including the random selection strategy.

We wish to point out that metamorphic testing is to be used with other test case selection strategies. Based on each test case selected according to a specific strategy and its output result, a number of new test cases can be designed to further test the software and increase our confidence in the software. Furthermore, metamorphic testing can also be combined with Blum's program checker which stops after the verification of the corresponding output is done but does not suggest further testing. There is a practical consideration in Blum's checker. The execution time taken by the checker must be less than that taken by the program being checked. Checkers satisfying this constraint are said to have the *oh property* ([2]) Otherwise, the checker may be impractical. In some cases, under this time constraint, an affirmative answer of whether the output is correct may not be possible. Sorting is one of the examples that Blum used to illustrate this situation. In order to have an affirmative answer of whether the output is correct, $O(n \log n)$ time is required but it is too expensive. Randomized checking can be used to give a probabilistic confidence on the correctness of the result. Our method, on the other hand, can still be applied even when it is computationally expensive to do the checking. Like program checkers, metamorphic testing can be applied in both the testing and production phases.

As metamorphic testing generally requires the use of problem domain knowledge, the approach is illustrated using several program examples. Section 2 states the preliminaries and examples are presented in Section 3. Concluding remarks are presented in Section 4.

## 2 Preliminaries

Given a program $P$ and a test case $\overline{x}$, the corresponding output is denoted by $P(\overline{x})$. Assume $\overline{x_0}$ is a test case with output $P(\overline{x_0})$. Our approach is to construct a number of new test cases $\overline{x_1}$, $\overline{x_2}$, …, $\overline{x_k}$ based on the input-output pair (that is, $(\overline{x_0}, P(\overline{x_0}))$) and the errors usually associated with $P$. These new test cases are designed to reveal the errors which go undetected by the test case $\overline{x_0}$.

The time for the construction of each new test case and the checking on the corresponding output are assumed to be strictly less than the execution time of the program. This requirement is in line with Blum's oh property of program checkers. In practical situations, we should try to minimize the time for the construction and checking. Access to individual elements in the database or the data structure is allowed as long as the overall construction time and the checking time satisfy the above constraint. For example, if the program searches for a target in an array $A$, an inspection of $A[i]$ is allowed to design the next test case. There is another practical concern of the newly constructed test cases. In the production phase, these test cases should not modify the production database. However, in the testing phase, this is allowed. In other words, test cases which modify the content of databases should not be performed in the production phase. In section 3, a number of examples are given to illustrate this idea.

# 3 Examples

## 3.1 Example 1: Binary Search on Sorted Array

Consider a program which locates the position of a given key in a sorted array with distinct elements using binary search. The program should return the position of the key if it exists or a value of -1 otherwise.

Let us assume that the input is $(x, A[i..j])$ and the output is $k$. Depending on the input-output pair, we have the following cases:

*Case 1*: Suppose $A[k] \neq x$. Obviously, there is an error in the program.

*Case 2*: Suppose $k$ = -1. Either $x$ does not exist in $A$ or there is a program bug that makes the program report non-existence. If $A$ is known, this can be verified easily. However, under a production environment, $A$ is usually a huge database and it is impractical to scan the entire database in order to confirm that $x$ does not exist in $A$. To increase the confidence that the program does not commit this type of error, randomly select an entry, $A[p]$, and rerun the program using the test case $(A[p], A[i..j])$ for which the expected output is $p$. This can be considered as program checking. As a remark, in subsequent discussion, some of the test cases are useful in both testing and production environments while others may be more appropriate in only one of them.

*Case 3*: Suppose $A[k] = x$. Under the assumption that the program still has bugs despite this apparently correct output, there are at least two possible errors.

The "correct" output is due to a program bug which overwrites the content of $A[k]$ with $x$. Before designing a test case to detect this error, check $A[k-1] < x < A[k+1]$. If this is not correct, an error is revealed. Replace $A[k-1]$ and $A[k+1]$ with $-\infty$ and $+\infty$ respectively if either of them does not exist. Otherwise, choose a $y$ such that $A[k-1] < y \neq x < A[k+1]$. Rerun the program with the test case $(y, A[i..j])$ for which the expected value is -1. If it is not feasible to get a $y$ within this range, then try to choose a $y$ such that $A[k-r] < y \neq z < A[k+r]$ where $z \neq A[j]$ for $k-r < j < k+r$ from $r = 2$ onwards until either such a $y$ is found, or it is not feasible to find such a $y$, or a reasonable number of unsuccessful attempts have been tried. Similarly, if either $A[k-r]$ or $A[k+r]$ does not exist, replace it by $-\infty$ and $+\infty$ respectively.

The output is correct but there is a splitting error which is not revealed by this test case. Rerun the program using test cases, $(A[k-1], A[i..j])$ and $(A[k+1], A[i..j])$, which aim at uncovering this type of error. The expected results are $k-1$ and $k+1$ respectively. If either $A[k-1]$ or $A[k+1]$ does not exist, replace it by $A[\lfloor \frac{i+j}{2} \rfloor - 1]$ and $A[\lfloor \frac{i+j}{2} \rfloor + 1]$ respectively. The reason for choosing $(A[\lfloor \frac{i+j}{2} \rfloor - 1], A[i..j])$ and $(A[\lfloor \frac{i+j}{2} \rfloor + 1], A[i..j])$ as test cases is that $A[\lfloor \frac{i+j}{2} \rfloor]$ is guaranteed to exist and if there is a splitting error, both of these test cases have a similar capability as $A[k-1]$ and $A[k+1]$ in revealing this bug. If both of them do not exist, $A$ has only one element. Hence, no more testing is required. As an illustration, the following shows an incorrect function with a splitting error at line 13. The parameter passed to the recursive call is incorrect, the element $A[mid+1]$ is missing.

```
1.   int BinSearch(x,A,i,j)
2.   {
3.     if (i > j)
4.        return -1
```

```
5.    else
6.       {
7.          mid = floor((i+j)/2);
8.          if (A[mid]==x)
9.             return mid;
10.         if (A[mid] > x)
11.            BinSearch(x,A,i,mid-1);
12.         else
13.            BinSearch(x,A,mid+2,j); /* correct statement is:    */
14.      }                             /* BinSearch(x,A,mid+1,j); */
15. }
```

Let the input array $A$ be 4, 6, 10, 15, 18, 25, 40 and the key $x$ be 25, the function call BinSearch($x$, $A$, 1, 7) will return a value of 6. The output is correct with respect to this particular test case. Using the above constructed test cases, both the test cases (18, $A$, 1, 7) and (40, $A$, 1, 7) will reveal this error.

Table 1 summarizes these possible cases with the exceptional handling left out. Since the purpose of the examples is to illustrate the idea, the errors listed in each example are by no means exhaustive. Also, the array $A$ used in the examples is for illustration purposes only, it can represent a production database and need not be passed to the function by parameter.

| | New Test Case | | |
|---|---|---|---|
| Condition | Inputs | Suspected Error | Expected Result |
| $A[k] \neq x$ | None | Incorrect position returned | N.A. |
| $k = -1$ | $(A[p], A[i..j])$ for any $p$ | Report non-existence even if the key exists | return $p$ |
| $x > A[k+1]$ or $x < A[k-1]$ | None | Overwriting error | N.A. |
| $A[k-1] < A[k] = x < A[k+1]$ | $(y, A[i..j])$ where $A[k-1] < y \neq x < A[k+1]$ | Overwriting error | return -1 |
| | $(A[k-1], A[i..j])$ and $(A[k+1], A[i..j])$ | Splitting error | return $k-1$ and $k+1$ respectively |

Table 1: Possible Test Cases for Example 1

## 3.2 Example 2: $k$th Occurrence of $x$ in Unsorted Array

Given an input $(x, k, A[i..j])$, the problem is to locate the $k$th occurrence of a key $x$ from an unsorted array $A[i..j]$ where $j - i + 1 \geq k \geq 1$. The output should be the position of the $k$th occurrence of $x$ if it exists or -1 otherwise.

Assuming that the output is $p$, consider the following cases:

*Case 1*: Suppose $A[p] \neq x$. A program error is revealed.

*Case 2*: Suppose $p = -1$. To increase our confidence that the program will not report non-existence if the key does exist, randomly choose an element $A[r]$ and rerun the program with the test case $(A[r], 1, A[i..j])$. Since $A[r]$ is in the array, the expected result must be any valid index other than -1 and $p \leq r$.

*Case 3*: Suppose $A[p] = x$. That is, the output looks correct. There may be at least two possible errors.

One possible error is due to overwriting. The content of $A[p]$ may be overwritten by $x$, or the program starts overwriting the subsequent array entries only after it hits the first occurrence of $x$. For the former case, design two new test cases as follows. One is $(y, 1, A[p..p])$ and the other is $(y, 1, A[p-1..p+1])$ where $y \neq A[p-1], x, or A[p+1]$. If either $A[p-1]$ or $A[p+1]$ does not exist, ignore it when choosing the value of $y$ and change the second test case to $(y, 1, A[p..p+1])$ or $(y, 1, A[p-1..p])$ respectively. The expected results for both test cases are -1. The reason we need the second test case is that some of the loops inside the program may not be executed if $A$ has only one element. For the latter case, scan $A$ starting from the element $A[p]$ until an element $A[r] \neq x$. Rerun the program with the test case $(x, 2, A[r-1..j])$ for which the expected value is either greater than $r$ or is -1. However, if the scanning process takes too long or such an $r$ cannot be found (that is, $A[q] = x$ for all $p \leq q \leq j$), the construction of such a test case should be abandoned since the construction becomes impractical.

If there is no overwriting error, the program may report the $q$th occurrence of $x$ where $q \neq k$. The two test cases, $(x, 1, A[p..p])$ and $(x, 1, A[p..j])$, are designed to reveal this error. The expected results for both cases should be $p$.

Table 2 summarizes these possible cases.

| Condition | New Test Case | | |
|---|---|---|---|
| | Inputs | Suspected Error | Expected Result |
| $A[p] \neq x$ | None | Incorrect position returned | N.A. |
| $p = $ -1 | $(A[r], 1, A[i..j])$ for any $r$ | Report nonexistence despite the key exists | return $m$ $m \leq r$ |
| $A[p] = x$ | $(x, 1, A[p..j])$ or $(x, 1, A[p..p])$ | Report $q$th occurrence where $q \neq k$ | return $p$ |
| | $(y, 1, A[p..p])$ or $(y, k, A[p-1..p+1])$ where $y \neq A[p-1], x, A[p+1]$ | Overwriting error | return -1 |
| | $(x, 2, A[r-1..j])$[1] | | return -1 or $m$ where $m > r$ |

Table 2: Possible Test Cases for Example 2

The function given below has an initialization error at line 3. The variable $m$ should be initialized to 1, so the position returned by the function is, in fact, the $(k+1)$th occurrence of $x$. Under the test case (1, 2, 1, 3, 1, 4, 1, 3, 2), the output is 5. With the test case (1, 1, 1, 3, 2), the error can be revealed.

```
1.  int search(x,k,A,i,j)
2.  {
3.      int p=i, m=0;  /* correct initialization: m=1; */
4.      int found=0;
5.      while ((p<=j) && (!found) && (m<=k))
6.      {
7.          if (A[p] == x)
8.              m++;
9.          if (m==k)
```

---

[1] Refer to the description for how to determine $r$.

```
10.            found=1;
11.        else
12.            p++;
13.    }
14.    if (found)
15.        return p-1;
16.    else
17.        return -1;
18. }
```

## 3.3 Example 3: Shortest Path in an Undirected Graph

Given a weighted graph $G$, a source node $x$, and the destination node $y$ in $G$, the problem is to output the shortest path and the shortest distance from $x$ to $y$.

Let us assume that the output path is $x, v_1, v_2, \ldots, v_k, y$ and the corresponding distance is $p$. This output is difficult to check even in the testing phase if the input graph $G$ is non-trivial. The only trivial checking is to verify that $(x, v_1)$, $(v_1, v_2)$, ..., $(v_k - 1, y)$ are edges of $G$. In this case, a practically feasible test oracle does not exist. One common error of the program is that in extending the shortest path from $x$ to $y$, the minimum distances from $x$ to the unvisited vertices are not correctly updated. In other words, the subpath of the returned "shortest" path is not the shortest. Based on this particular fault, one of the possible test cases is $(y, x, G)$ for which the expected shortest distance is $p$. The rationale for this test case is that if the returned path is not the shortest due to the updating error for the partial path, the program error will most likely give rise to a different path of different length as different updating error may be made. Clearly, this test case is not applicable in a directed graph. Another possible test case with the same design principle, which can be used in both undirected and directed graphs, is $(x, v_i, G)$ and $(v_i, y, G)$ where $1 \leq i \leq k$. Other possible test cases based on this particular fault are shown in Table 3.

| Condition | New Test Case | |
|---|---|---|
| | Inputs | Expected Result |
| $x, v_1, v_2, \ldots, v_k, y$ is a feasilbe path | $(y, x, G)$[1] | return the same shortest distance, $p$. (path may not be the same) |
| | $(x, v_i, G)$ and $(v_i, y, G)$ for any $1 \leq i \leq k$ | the sum of two distances $= p$ |
| | $(y, v_i, G)$[1] and $(v_i, x, G)$[1] for any $1 \leq i \leq k$ | |
| | $(v_1, v_k, G)$ | the sum of returned distance and length of $(x, v_1)$ and $(v_k, y) = p$ |
| | $(v_k, v_1, G)$[1] | |

Table 3: Possible Test Cases for Example 3

The following function should output the shortest path and calculate the shortest distance from $x$ to $y$ in the graph $G$. However, there is an updating error in the statement at line 13. The correct statement should be D[v] = min(D[v], D[k]+G[k,v]).

---

[1] These test cases are not applicable if $G$ is a directed graph.

```
1.   int ShortestDist(x,y,G)
2.   {
3.     S={x}; for all i, P[i]=x;    /* store shortest path */
4.     int i, k=0;
5.     for (i=1; i<=n; i++)
6.         D[i]=G[x][i];
7.     while (k != y)
8.     {
9.        choose k in V-S
10.           such that D[k] is minimum;
11.       S=S+{k};
12.       for each vertex v in V-S    /* correct statement:          */
13.           D[v] = D[k] + G[k,v];   /* D[v]=min(D[v],D[k]+G[k,v]); */
14.           if (D[k]+G[k,v]) <= D[v]
15.               P[v] = k;           /* update path */
16.       return D[k];
17.  }
```

Consider the following graph $G$. If the input is $(c, a, G)$, then the output will be 34 and the shortest path will be *cdba*. If $G$ is large, it is difficult to verify the answer. However the test case $(a, c, G)$ and the test cases $(c, b, G)$ and $(b, a, G)$ together are able to reveal this program error.

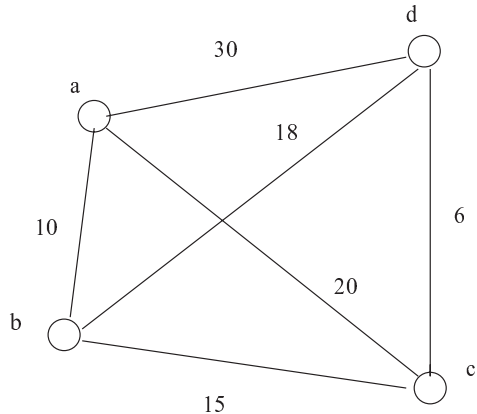

Figure 1: An Input Graph for Example 3

## 3.4 Example 4: Solving a System of Linear Equations by Gaussian Elimination

Consider a program which solves a system of linear equations, $Ax = b$ by using Gaussian elimination where $A$ is the square matrix of coefficients and $b$ is a column vector. The idea of Gaussian elimination is to perform elementary operations on the augmented matrix $[A|b]$ until either the resulting matrix $[I|c]$ is found where $I$ is the identity matrix and $c$ will be the solution, or the procedure returns no solution if matrix $A$ is singular.

Let us assume that the input is $(A, b)$ and the output is $x_0$. To check whether $x_0$ is a solution of the system, we can subsitute $x_0$ into the equations. This can be done manually if the number of equations is small or using another program. This approach can be easily adopted in the program checker. Depending on the input-output pair, we have the following cases:

*Case 1*: Suppose $Ax_0 \neq b$. Obviously, there is an error in the program.

*Case 2*: Suppose $Ax_0 = b$. Under the assumption that the program still has errors despite this apparently correct output, there are at least three possible types of errors in the program according to the domain knowledge that this type of programs will usually contain three main parts in each step of the elimination: choosing an appropriate pivot, performing elementary operations to transform $A$ to $I$, and performing elementary operations to transform $b$ to $c$. In this paper, only the first type of errors will be discussed. A comprehensive treatment of the detection of other errors in numerical problems using metamorphic testing can be found in [3].

The idea for using metamorphic testing comes from the fact that if we prepare a new input pair by interchanging any two rows of $A$ and the corresponding entries of $b$, the output will be the same as the output given the original input pair $(A, b)$. The Guassian elimination involves the selection of a pivot. If there are errors in the program, it may select for the new input pair a different pivot or make another mistake when transforming $(A, b)$ to $(I, c)$. As a result, the program gives a different output for the new input pair. Similarly, the input pair for additional test cases may be generated by interchanging the columns of $A$ to derive new test cases. To be more precise, let the original input pair be $(A, b)$ and the corresponding output be $x$ where $x = (x_1, x_2, \ldots, x_n)$. Upon a successful test case, two additional test cases can be constructed. First, we can pick two rows arbitrarily in the input pair, say row $i$ and row $j$. Interchange entries of row $i$ with that of row $j$ to give a new matrix $A'$. Swap the $i$th entry of $b$ with the $j$th entry to give $b'$. The expected output $x'$ of the new input pair $(A', b')$ should be equal to $x$. Otherwise, an error is revealed. Similarly, we can construct another test case by interchanging two columns in $A$, say column $i$ and column $j$, to give a new matrix $A''$. Using the new input pair $(A'', b)$, the expected output will be $x''$ where $x''_i = x_j$, $x''_j = x_i$, and $x''_k = x_k$ for $k \neq i, j$. Table 4 summarizes these two additional test cases.

| Condition | New Test Case | |
|---|---|---|
| | Inputs | Expected Result |
| $Ax = b$ | $(A', b')$ where $A'$ is obtained from $A$ by swapping row $i$ and $j$ and $b'$ is obtained from $b$ by swapping corresponding entries. | $x' = x$ |
| | $(A'', b)$ where $A''$ is obtained from $A$ by swapping col $i$ and $j$ with $i < j$. | $x'' = (x_1, \ldots, x_{i-1}, x_j, x_{i+1}, \ldots, x_{j-1}, x_i, x_{j+1}, \ldots, x_n)$ |

Table 4: Additional Test Cases for Example 4

*Case 3*: $A$ is singular. This case will not be discussed in this paper. Please refer to [3] for more details.

As an example, consider the following function. It contains an initialization error in variable *max* at line 5 which will lead to an incorrect selection of pivot. Let the input pair $(A, b)$ be:

$$A = \begin{bmatrix} 1 & 2 & 3 \\ 2 & 2 & 3 \\ 3 & 3 & 3 \end{bmatrix}$$

$$b = \begin{bmatrix} 1 \\ 1 \\ 1 \end{bmatrix}$$

The output $x = (0, 0, \frac{1}{3})$ is correct. However, if we interchange row 2 with row 3 in $A$, the program will report no solution. It should be pointed out that if the program checker verifies the output only by subsituting the vector $x$ back to the equations, this error may not be detected.

```
1.   void Gauss(A,b)
2.   {
3.      int max;            /* used for locating pivot */
4.      int pivot = 0;
5.      for j = 1 to n {    /* transformation of [A|b] to [I|c] */
6.         max = 2;         /* correct initialization: max = 0 */
7.         for i = j to n   /* locating pivot */
8.            if (abs(A[i][j]) >= max) {   /* update pivot */
9.               max = abs(A[i][j]);
10.              pivot = i; }
11.        if (pivot <> j) {              /* assume no error */
12.           interchange(A[pivot], A[j]);
13.           interchange(b[pivot], b[j]); }
14.        if (A[j][j] == 0)
15.           return(-1);                 /* no solution */
16.        remaining steps   }            /* assume no error */
17.     Calculate solution vector x; /* [I|c] should be obtained */
18.  }
```

## 4 Conclusion

In this paper, we have proposed a new approach for generating test cases called *Metamorphic Testing*. This approach aims at uncovering those errors which are commonly found in similar applications. Test cases are generated according to the input-output pair of an (apparently) successful test case, which is selected using some well known criteria. The test cases generated target those errors which are possibly left undetected by the previous successful test case. Metamorphic Testing enables software practitioners to design additional tests and hence improve the software reliability. It augments the effectiveness of the existing testing strategies by making use of the corresponding successful test cases to construct additional test cases.

In our approach, the design of next test cases *need* not depend on complicated theories. Rather it is based on programming experience and testing experience. Examples are given to show that this approach is feasible and that the newly constructed test cases are useful in revealing some common errors. The approach is applicable in both the testing and production phases and does not rely on the assumption of a test oracle. We find its application to the situations without test oracles particularly interesting as the problem is legitimate and has not been adequately addressed by existing testing techniques. Another advantage of the proposed approach is that it does not depend on any particular test case selection strategy, so it can be easily used along with other test case selection strategies and can be easily combined with Blum's program checker.

While the primary objective of this paper is to present the motivation of the approach and its applicability, a methodology supporting such an approach has yet been fully developed. Our ultimate goal is to provide a concrete methodology leading to the pragmatic construction of "next" test cases. As metamorphic testing requires problem domain knowledge in general, we expect that a domain specific methodology should be developed. To facilitate the development of such a methodology, we are studying various criteria of classifying programs into their appropriate domains.

**Acknowledgements**

We would like to thank F.T. Chan and M.F. Lau for helpful discussion on the paper. In particular, example 4 is extracted and modified from an example in a manuscript by Chan, Lau, and the authors..